# The growth of bismuth on $Bi_2Se_3$ and the stability of the first bilayer

Haoshan Zhu, Weimin Zhou and Jory A. Yarmoff*

*Department of Physics and Astronomy, University of California, Riverside, Riverside, CA 92521*

**Abstract**

Bi(0001) films with thicknesses up to several bilayers (BLs) are grown on Se-terminated $Bi_2Se_3$(0001) surfaces, and low energy electron diffraction (LEED), low energy ion scattering (LEIS) and atomic force microscopy (AFM) are used to investigate the surface composition, topography and atomic structure. For a single deposited Bi BL, the lattice constant matches that of the substrate and the Bi atoms adjacent to the uppermost Se atoms are located at fcc-like sites. When a 2nd Bi bilayer is deposited, it is incommensurate with the substrate. As the thickness of the deposited Bi film increases further, the lattice parameter evolves to that of bulk Bi(0001). After annealing a multiple BL film at 120°C, the first commensurate Bi BL remains intact, but the additional BLs aggregate to form thicker islands of Bi. These results show that a single Bi BL on $Bi_2Se_3$ is a particularly stable structure. After annealing to 490°C, all of the excess Bi desorbs and the Se-terminated $Bi_2Se_3$ surface is restored.

---

*Corresponding author, E-mail: yarmoff@ucr.edu

**I. Introduction**

Topological Insulators (TIs) are promising materials for superconductor, spintronics and quantum computing applications because of their intrinsic topological surface states (TSS) that are protected by time-reversal symmetry [1-3]. $Bi_2Se_3$ is one of the most well-known TIs due to its comparatively large, and thus practical, bulk band gap of 0.35 eV [2] and it's simple surface band structure in which the TSS form a Dirac cone [4]. $Bi_2Se_3$ has a rhombohedral structure belonging to the $R\bar{3}m$ space group. Five two-dimensional (2D) hexagonal lattices of Bi and Se are stacked together along the [0001] direction in the sequence of Se-Bi-Se-Bi-Se to form a quintuple layer (QL). Adjacent QLs are connected to each other by weak van der Waals forces. When a high quality pure $Bi_2Se_3$ single crystal is cleaved in ultra-high vacuum (UHV) or subjected to ion bombardment and annealing (IBA) in UHV, the resulting surface is highly ordered and terminated by Se at the top of a complete QL [5].

Bulk Bi is also a 2D material that is composed of bilayers (BLs) bonded to each other by van der Waals forces [6]. A single Bi(0001) BL is also reported to be a 2D topological insulator [7,8]. Ultrathin Bi films of only a few BLs are attracting more interest for their nontrivial properties such as a large magnetoresistance and topological edge states [9,10]. Hetero-structure engineering with Bi BLs and other materials, including 3D TIs, are becoming popular because of the ability to control their electronic properties and fabricate unique device structures [11-13]. Recent studies show giant Rashba-split states in Bi BL-terminated $Bi_2Se_3$ surfaces prepared by both epitaxial growth [14-16] and hydrogen etching methods [17,18].

Deposited Bi grows on TI surfaces at room temperature along the [0001] direction in a quasi bilayer-by-bilayer mode, and has the same hexagonal lattice symmetry as $Bi_2Se_3$ [19,20]. Previous STM work shows that the first Bi bilayer is strongly compressed on $Bi_2Se_3$(0001) substrate so that it matches the lattice parameter of the substrate. Further Bi deposition forms BL islands that eventually coalesce to complete bilayers with sufficient coverage. A periodic buckling of the lattice is found beginning from deposition of the $2^{nd}$ BL and lasting until the $5^{th}$ BL is deposited. The Bi BL lattice gradually relaxes to become bulk-like at a coverage of 24 BLs.

An interesting aspect of the Bi/$Bi_2Se_3$ system is the unique stability of a single deposited BL. Density functional theory (DFT) calculations show that the single BL-terminated structure has a lower surface energy than the bare Se-terminated surface [21,22] and there is a stronger bonding between a QL and a Bi BL than there is between QLs [23,24]. In support of this notion, $Bi_2Se_3$ surfaces terminated with a single Bi BL have been found to spontaneously form on *in-situ* [22] and *ex-situ* [25] cleaved samples. In addition, there is a report that exposing a Se-terminated surface to air for 5 minutes can form a stable Bi BL on the surface, presumably through a (still-unidentified) surface chemical reaction [26].

The present paper presents a low energy electron diffraction (LEED), low energy ion scattering (LEIS) and atomic force microscopy (AFM) study of Bi BLs grown on $Bi_2Se_3$(0001) by molecular beam epitaxy (MBE). It is confirmed that the first deposited BL grows epitaxially, and it is further shown that the Bi atoms in this first BL sit in fcc-like sites. As shown previously, subsequent layers grow as islands, since they are affected by the strain resulting from the differences in the lattice constants of $Bi_2Se_3$ and Bi, until the film is thick enough to form what is essentially bulk single crystal Bi. A relatively low temperature annealing of the thicker films shows that the first BL remains in place as the rest of the deposited Bi coalesces into islands, which is another sign of the unique stability of the $1^{st}$ BL.

**II. Experimental procedure**

Single crystal $Bi_2Se_3$(0001) is used as the substrate. Bulk $Bi_2Se_3$ was prepared by melting stoichiometric mixtures of Bi (99.999%, Alfa Aesar) and Se shot (99.999+%, Alfa Aesar) in an evacuated quartz ampule ($2\times10^{-6}$ Torr) with an inner diameter of 17 mm, and then following a slow-cooling procedure [5]. Approximately 1 cm × 1 cm × 2 mm single crystal $Bi_2Se_3$ plaques are cleaved from the bulk $Bi_2Se_3$ crystals and attached to a transferable tantalum (Ta) sample holder (Thermionics) using spot-welded Ta strips.

Most of the measurements are performed in an ultra-high vacuum (UHV) main chamber that has a base pressure better than $2\times10^{-10}$ Torr. There is a load-lock chamber that enables the sample holders to be inserted without the need to vent and re-bake the main chamber. The samples are transferred onto an x-y-z manipulator that allows for rotation about both the polar and azimuthal axes and includes an electron-beam heater. For the measurements performed here, the samples are heated radiatively without applying a bias voltage to the filament, as the annealing temperature needed to prepare $Bi_2Se_3$ is rather low. This main chamber contains a sputter gun (Physical Electronics) for sample cleaning, optics for LEED (Princeton Research), and the equipment needed for LEIS that is described below. The sample is held at room temperature during the collection of LEED images and LEIS spectra.

A Se-terminated $Bi_2Se_3$(0001) surface is prepared in the main UHV chamber by ion bombardment and annealing (IBA), which involves 30 min of 500 eV $Ar^+$ ion bombardment at a current density of $2.5\times10^{12}$ $cm^{-2}$ $sec^{-1}$ with the sample at room temperature followed by annealing at 490°C for 30 min. This IBA procedure produces high quality clean and well-ordered surfaces, as described elsewhere [27]. The annealing temperature is calibrated by a thermocouple attached to the Ta sample holder, but the actual temperature of the surface can vary from -50°C to +20°C from the reported value as the thermocouples are not attached directly to the samples. Also, thicker samples require more annealing time to allow the surface to reach the desired temperature.

Bi(0001) BLs are grown on $Bi_2Se_3$(0001) surfaces at room temperature using a molecular beam epitaxy (MBE) system that is attached to the main chamber such that samples can be transferred under UHV. Bi is evaporated at a rate of 1.45 Å min$^{-1}$, as calibrated by a quartz crystal microbalance (QCM), from a Knudsen cell heated to 530°C. The amount of Bi deposited is also confirmed using AFM images, as discussed below.

The rear-view LEED system (Princeton Research Instruments) is used to ascertain the sample cleanness, crystallinity and orientation of the surface unit cell, and to monitor how the lattice parameter changes with the growth of Bi films. The sample position is kept the same with respect to the LEED optics for each measurement. The electron beam is normally incident onto the surface with the beam energy fixed at 21.8 eV.

Low energy ion scattering (LEIS) time-of-flight (TOF) spectra are used to identify the surface elemental composition [28] and measure the neutral fraction (NF) of scattered $Na^+$ [29]. A 3.0 keV $Na^+$ ion beam (Kimball Physics IGS-4) pulsed at 100 kHz is normally incident onto the sample surface and the scattered projectiles are collected at a scattering angle of 125° by a microchannel plate (MCP) detector located at the end of a 0.57 m long flight tube. A bias voltage of 400 V is applied to deflection plates in the flight tube to separate scattered ions from neutrals, allowing for independent collection of spectra for the total scattered yield and the scattered neutrals. The bias voltage is periodically turned on and off every 60 sec while both spectra are collected simultaneously to avoid any effects of long term drift in the incident ion beam current. The front of the MCP is held at ground potential so that the scattered neutrals and ions impact the detector with the same kinetic energy.

Impact collision ion scattering spectrometry (ICISS) is used to probe the surface atomic structure [30]. ICISS is performed using a 3.0 keV Na$^+$ ion beam and a Comstock electrostatic analyzer (ESA), which measures only scattered ions, by fixing the scattering angle at 161° and rotating the sample about the polar axis. An energy spectrum is first collected to locate the positions of the Bi single scattering peak (SSP) and the Se SSP. The detection energy is then fixed at one SSP energy and the intensity of that SSP is monitored with respect to the incident polar angle as the sample is rotated. The energy is then set to the other SSP energy, and the ICISS data collection procedure is repeated. The sample manipulator has a computer-controlled stepper motor that automatically rotates the sample to enable quick and reproducible collection of the ICISS angular distributions.

AFM images are collected using a separate Dimension 3000 (Digital Instruments) apparatus. The samples used for AFM are cleaved *in situ* under UHV, as this produces larger terraces that are better suited for AFM than IBA-prepared samples [25,31]. Bi is deposited in UHV using the MBE system, and the samples are then removed from vacuum and transported to the AFM instrument. The tapping mode images are collected in air at room temperature using TESPA-V2 silicon tips (Bruker).

**III. Results**

AFM measurements are conducted to both calibrate the coverage of the Bi BLs and to monitor changes of the surface topography. Figure 1 shows AFM images collected after various treatments. Figure 1(a) shows the surface of Bi$_2$Se$_3$ following *in situ* cleaving in UHV, although the sample was removed from vacuum to collect the image. The IBA-prepared samples are essentially atomically flat, but they contain QL-high steps that separate terraces with widths of approximately 200 nm [27]. In contrast, *in-situ* cleaved samples contain wider terraces, which makes them more suitable for calibrating the Bi coverage with AFM. Figures 1(b)-(e) show images collected from Bi films grown on $Bi_2Se_3$. The green curve labeled A in Fig. 1(g) shows the profile of the line shown in the 1.0 BL image (Fig. 1(c)). The heights of the features seen in Fig. 1(g) are all about 0.5 nm, which corresponds to that of a single Bi BL, and this value is indicative of all of the features observed in the AFM images. The Bi coverage is calculated from the ratios of the accumulated area of the Bi islands or films to the area of the entire substrate. Analysis of the areas measured for Bi evaporation times of 2.5, 3.5 and 5.5 min indicate that these substrates are covered with 0.7, 1.0 and 1.5 BL, respectively. This implies that the Bi evaporation rate is about 1.4 Å $min^{-1}$ on average, which is consistent with the value determined using the QCM (1.45 Å $min^{-1}$). A 6 Bi BL covered surface is also examined, as shown in Fig. 1(e), in which the surface is fairly flat with a height variance of $\pm 0.5$ nm, as shown by line profile B in Fig. 1(g), but it still has features with heights that are integral numbers of Bi BLs. The AFM images confirm that Bi grows in quasi bilayer-by-bilayer mode at high coverages and there's no evidence of tall islands for a 6 BL coverage at room temperature.

LEED is used to monitor the surface symmetry and determine how the lattice parameter changes with Bi deposition. Figure 2 shows LEED patterns collected for different coverages of Bi. Note that the LEED image and background appear dimmer on the right half of the screen due to a degradation of the screen coating. A pristine Se-terminated $Bi_2Se_3$ substrate displays a 1×1 hexagonal pattern, as seen in Fig. 2(a). Three-fold LEED spots that are observed at higher beam energies are not shown here. When a single Bi BL is deposited onto the substrate, the LEED spots shift slightly towards the center and get much brighter, as seen in Fig. 2(b) [16]. Three-fold higher-order satellite spots start to show up at 1.7 BL, while the spots from the 1st BL are still visible, which indicates that a Moiré structure is formed on substrate. The satellite spots get clearer and brighter when the Bi coverage reaches 3 BLs at the expense of decreasing the intensity of the LEED spots associated with the 1st Bi BL. The satellite spots only appear if the $Bi_2Se_3$ substrate and the Bi film are both clean and well-ordered. For example, if the annealing temperature is too low or the annealing time too short during the IBA process, then the LEED spots are blurry. The satellite spots gradually fade out as more Bi is deposited, while the LEED spots corresponding to bulk Bi(0001) get brighter (Figs. 2(d)-(f)). The LEED pattern eventually evolves to that of a bulk-like Bi(0001) single crystal (Fig. 2(g)) when the coverage reaches about 8 BLs or more.

Based on the average of the measured distances between the centers of equivalent LEED spots (such as line D in Fig. 2(b)) and using the known $Bi_2Se_3$ surface lattice constant of 4.14 Å [32] for calibration, the lattice constants of the outermost Bi BLs are calculated, which are shown in Fig. 3 as a function of Bi BL coverage. The lattice parameters shown in Fig. 3 have error limits of ±0.05 Å as estimated from the standard deviation of multiple measurements. These data are reasonably consistent with the results of Ref. [19], although they show a less gradual change of lattice constant with Bi film thickness. Although this can be due to the finer gradations of coverage used in Ref. [19] combined with the lower resolution of our LEED optics, another possibility is that LEED probes deeper than RHEED and is thus less sensitive to changes in the uppermost layer lattice constant. Thus, by the time that LEED shows a change in lattice constant, it is probing multiple bilayers.

The average grain size of the first bilayer can be estimated from the LEED spot width and the distance between the LEED spots. The measured width of LEED spots is the sum in quadrature of the intrinsic resolution of the LEED optics ($\sigma_{intrinsic}$) and the broadening caused by a limited surface domain size ($\sigma_{grain}$). The measured LEED spot width is 2.3% of the Brillouin zone (BZ) for an IBA prepared Bi$_2$Se$_3$ substrate, and 5.6% of the BZ for a 1 Bi BL covered sample. The intrinsic width, $\sigma_{intrinsic}$, is unknown, but a range of possible grain sizes can be estimated. By assuming that $\sigma_{intrinsic} = 0$, the lower limit of the grain size is the inverse of the width due to the finite grain size, which is equal to $\frac{1}{\sigma_{grain}} = \frac{1}{\sigma_{measured}} = \frac{1}{5.6\%} = 17.8$ atoms. By assuming that the Bi$_2$Se$_3$ substrate is perfect with an infinite grain size, the width measured from the clean substrate is then equal to $\sigma_{intrinsic}$ and the upper limit of the 1$^{st}$ BL grain size is calculated to be $\frac{1}{\sigma_{grain}} = \frac{1}{\sqrt{\sigma_{measured}^2 - \sigma_{intrinsic}^2}} = \frac{1}{\sqrt{5.6\%^2 - 2.3\%^2}} = 19.6$ atoms. Thus, the estimated diameter of the grains in the first Bi BL is around 19 atoms.

The lattice constant of the superlattice formed by the 1$^{st}$ and 2$^{nd}$ Bi BLs can be calculated by inserting the lattice constants of the 1$^{st}$ and 2$^{nd}$ BLs into the Moiré equation,

$$d = \frac{a_1 \times a_2}{|a_1 - a_2|},$$

where $a_1$ is lattice constant of the 1$^{st}$ Bi BL (4.2 Å) and $a_2$ is the lattice constant of the 2$^{nd}$ Bi BL (4.48 Å). From this analysis, it is found that $d$ = 67 Å, which suggests that the two lattices line up such that 15 unit cells of the 2$^{nd}$ Bi BL line up with 16 unit cells of the substrate. As the Bi coverage increases beyond 2 BLs, the lattice constant remains nearly the same, but expands very slightly until it reaches the Bi(0001) bulk value of 4.54 Å [6] at a thickness of about 5 BLs.

LEIS is a surface sensitive technique that is typically used to investigate the atomic composition and structure of the first few atomic layers of a solid [28]. Low energy ions (1 to 10 keV) only penetrate the first few atomic layers, and a double alignment scattering geometry can be chosen to make the spectra sensitive to only the outermost surface atoms [5,22,33]. In addition, when alkali ions are used as the projectiles, the probability that they are neutralized is related to the local electrostatic potential (LEP) just above the scattering site [34-36].

LEIS TOF spectra are used here to identify the surface elemental composition and measure the neutral fraction (NF). Figure 4 shows TOF spectra collected from $Bi_2Se_3$ samples with different Bi coverages. Note that the time scale in the figure is reversed so that increasing energy of the scattered projectiles is towards the right. The most prominent features in the spectra are the single scattering peaks (SSPs), which are labeled by the target atom symbol in the figure. A SSP corresponds to a projectile that experiences a single hard collision with a surface target atom and backscatters directly into the detector. The target atoms can be considered to be unbound atoms located at the lattice sites, as the kinetic energy of the projectiles is much greater than the bonding energy of atoms in a solid. Because low energy ions have very small de Broglie wavelengths, the energy lost in a single collision is primarily through a classical binary elastic collision so that the kinetic energy of the SSPs provide the mass distribution of the surface atoms. The peak at 4.1 μs in Fig. 4 is the Bi SSP, while the peak at 5.6 μs is the Se SSP. The data is analyzed by integrating each SSP after subtracting the background caused by multiple scattering, as described elsewhere [34]. This allows for a determination of the site-specific neutral fraction (NF) by calculating the ratio of the neutral yield to the total yield for each SSP [29]. Similarly, the ratio of the total yield Se SSP to the total yield Bi SSP can be determined, which is useful for monitoring the surface composition.

For the TOF spectra collected here, the incident ion beam is aligned normal to the sample surface and the azimuthal angle with respect to the crystal is adjusted. When the incoming projectiles are normal to the sample, only the outermost three atomic layers are directly visible to the beam for a bulk-terminated $Bi_2Se_3$ surface structure [5,22]. When the outgoing direction is parallel to the [10$\bar{1}$0] azimuth at a scattering angle of 125º (35° from the surface normal), as illustrated in Fig. 5(a), then both the incident and outgoing trajectories are along low index crystal lattice directions, which is a double alignment orientation. In this orientation, the projectiles scattered from the second and third layers are blocked from reaching the detector by the first layer atoms, as illustrated by the "blocking cone" in Fig. 5(b), so that the SSPs in the spectra represent only those atoms in the outermost layer. Maintaining normal incidence, but using other outgoing azimuthal angles, such as the $[01\bar{1}0]$ direction, produce single alignment orientations in which particles singly scattered from second and third layer atoms can also reach the detector.

In the double alignment TOF spectra collected from clean $Bi_2Se_3$ shown in the upper left panel of Fig. 4, the near absence of the Bi SSP verifies that the surface is Se-terminated [5]. The appearance of the Bi SSP in the single alignment orientation in the upper right panel of Fig. 4 shows that Bi is present in the second or third atomic layer. In both alignments shown in Fig. 4, the Se SSP decreases in intensity and the Bi SSP increases with Bi deposition, as expected. Note that the cross section for scattering 3.0 keV Na from Bi is ~2.3 times larger than for scattering from Se [37], which needs to be considered when analyzing the ratio of the Se and Bi SSPs.

The diamonds in Fig. 6 show how the Se:Bi ratio in single alignment varies with increasing Bi coverage, using the right y-axis. The ratio is obtained by dividing the areas of the SSPs and normalizing by the differential scattering cross sections [37] and MCP efficiencies at the kinetic energy of each SSP [38]. In single alignment, the beam can detect down to the third layer, so that Se SSP disappears after the 2$^{nd}$ Bi bilayer is deposited causing the Se:Bi ratio to become zero.

Figure 6 also shows how the Bi SSP NF in both alignments varies with increasing Bi coverage using the left y-axis. The NF in both orientations increases dramatically when the 1$^{st}$ Bi BL is deposited and increases more slowly until they both reach approximately 50% after 4 BLs are deposited. The increase of the NF is consistent with a more metallic surface being produced by the deposited Bi. According to the resonant charge transfer (RCT) model [35] that is normally used to describe the neutralization of scattered low energy alkali ions, the NF in scattering from a conventional solid increases as the local work function above the scattering site is reduced. Therefore, an increase of the neutral fraction with Bi deposition is expected, since the work function of Bi(0001) is 4.27 eV [39], which is lower than the 5.40 eV work function of $Bi_2Se_3$ [40].

The NFs in single and double alignment are different in scattering from the first Bi bilayer but become the same after the $2^{nd}$ BL is deposited. The difference in NFs when there is a single BL deposited shows that the chemical environments above the top and bottom Bi atoms in the first bilayer are different. In general, an upward dipole on surface decreases the local work function and thereby increases the measured NF [41,42]. The higher NF in double alignment than in single alignment indicates there are more electrons accumulated under the top Bi atoms in the first bilayer, which is consistent with the calculation in Ref. [43]. Such inhomogeneous distribution of charge can produce a strong electric field and result in the commonly observed Rashba effect [14-17]. When two or more BLs are deposited, the BLs are more freestanding and the chemical environment above the Bi atoms in each BL are then equivalent.

Impact collision ion scattering spectroscopy (ICISS) is performed to investigate the surface structure of the Bi bilayers. ICISS is a specialized version of LEIS in which the intensity of singly scattered ions is monitored as a function of the polar angle between the incident ion beam and the sample surface plane [30,44,45] Figure 5(b) illustrates the relationship between representative shadow cones and the atoms in the outermost two atomic layers. At very low incidence angles, i.e., close to the surface plane, each surface atom falls within the shadow cone of its neighbor so that no backscattering occurs. As the polar angle is increased, the edge of each surface atom shadow cone passes through its neighboring atom, causing ions to impact that atom and backscatter at the SSP energy. The ion flux at the edge of a shadow cone is enhanced relative to that of the incident beam due to small angle scattering. Thus, when the incident polar angle is continuously increased, a maximum appears in the scattering yield, called a flux peak, at the angle at which the edge of the shadow cone passes through the scattering atom. When the edge of a shadow cone formed by one first layer atoms hits the neighboring first layer atoms, as illustrated by leftmost shadow cone in Fig. 5(b), the peak is called a surface flux peak (SFP). Other flux peaks appear at larger angles when the shadow cone edge passes through a deeper lying atom.

The symbols in Figure 7 show separate ICISS angular scans for scattering from Bi and Se atoms that were collected from as-prepared $Bi_2Se_3$ (squares) and from that surface covered with 1 BL of Bi (circles). The clean $Bi_2Se_3$ surface shows a clear Se SFP at 10º, while there is little Bi SSP intensity in that same angular region, which is expected for a Se-terminated surface. When 1 Bi BL is deposited, the Se SFP attenuates and a Bi SFP grows, which indicates that the surface Se atoms are replaced by Bi. The shapes of Bi SSP scans in the large angle (20°-90°) region are nearly unchanged with Bi coverage, except for a slight enhancement of the peak at 55°, which indicates that Bi adlayer has the same basic atomic structure as a clean $Bi_2Se_3$ surface with Bi replacing Se.

Molecular dynamics (MD) simulations of ICISS are performed using the Kalypso software package [46] as done in previous work with $Bi_2Se_3$ [47]. The Thomas-Fermi-Molière repulsive potential using the Firsov screening length, corrected by a factor of 0.8, is used to calculate each projectile-target atom interaction. The acceptance angle used is 4° to match the experimental data. The target model is a two-dimensional ($1\bar{2}10$) plane as shown in the schematic diagram of Fig. 5(b). Periodic boundary conditions are applied parallel to the surface. The atomic positions at the $Bi_2Se_3$ surface are taken from the structural parameters determined by LEED in Ref. [48]. The vertical distance between the top two atomic layers is set to 1.55 Å. The height of the deposited Bi bilayer is set at 1.71 Å and distance from bottom of the bilayer to the top Se surface atoms is set at 1.96 Å to best fit the experimental data.

The upper panels in Fig. 7, which include the experimental data for the Se-terminated $Bi_2Se_3$ surface as square symbols, also show the simulations performed for the bulk structure as a solid line. Note that the simulations only reproduce single scattering events, while the experimental data contain a background of multiple scattering and out-of-plane scattering, which is more pronounced for the Se SSP scans because of the lower scattered energy, so that the background intensity of the experimental scans is generally larger than that of the simulations. What is important to focus on when analyzing these data, however, is the existence and positions of particular flux peak features since they indicate when the edge of a shadow cone impacts another atom, which is dependent on the specific crystal atomic surface structure. There is a good agreement between the positions and relative intensities of the flux peaks in the experimental data and simulations for the clean $Bi_2Se_3$ surface, which indicates the reliability of the simulations in reproducing angular scans for a given surface structure.

For the Bi BL-covered surface, three types of stacking orders are modeled to determine which one best fits the experimental data, which are shown as filled circles in both the upper and lower panels of Fig. 7. The simulations for the various stacking orders are separated into either the upper or lower panels to make it easier to distinguish them from each other. The first stacking order assumes that the Bi layer next to uppermost Se layer sits in fcc-like sites, the second places the Bi atoms in hcp-like sites, while the third considers the film to form a superlattice with the substrate. In an fcc-like stacking, the lower Bi atoms in the 1$^{st}$ BL are positioned directly above the 3$^{rd}$ layer Se, as is shown in Fig. 5. For hcp-like stacking, the lower Bi atoms in the 1$^{st}$ BL are positioned directly above the 2$^{nd}$ layer Bi. The first two models use the lattice parameter of the substrate (4.14 Å) as that of the Bi adlayer. The simulation of fcc stacking of the Bi bilayer, shown as green dashed lines in the top panels of Fig. 7, shares similar features with the experimental Se SSP scan, such as the positions of the rising edge at around 57° and the peak intensity at 81°, which are marked by vertical dashed lines. In addition, the Bi SSP scan shows an enhancement of the peak at 55° in the simulated scan, which is consistent with the experimental data. The hcp stacking simulation, shown by the black dashed lines in the bottom panels, has an extra feature in the Se SSP scan at around 40°, which is not observed in the experimental data. A superlattice, based on the different lattice parameters of $Bi_2Se_3$ and Bi, has 69 Bi atoms per single adlayer with a lattice constant of 4.2 Å and 70 Se atoms per substrate layer with a lattice constant of 4.14 Å, is used as the target in the simulation shown by the red dashed lines in the lower panels. This model can be considered as an average of the fcc, hcp and on-top stacking, which has the effect of suppressing all the features in the large angle (20°-90°) region and thus does not provide a good match to the measured scans. Thus, the fcc stacking model provides the best match to the experimental data.

Two different annealing temperatures are used to modify the surface composition and structure. Figure 8 shows the LEED patterns collected from an as-deposited 3 BL film and after annealing that sample at different temperatures, while Fig. 9 shows the corresponding TOF spectra collected in a double alignment orientation. Before annealing, the LEED in Fig. 8(a) shows the satellite spots and the upper panel in Fig. 9 shows primarily a Bi SSP, as all of the Se in the substrate is covered by the Bi film.

The lower annealing temperature of 120°C is used to investigate the connection between the $Bi_2Se_3$ surface and the 1st Bi BL. The LEED pattern in Fig. 8(b) suggests that the surface Bi crystalizes after annealing at 120°C, which makes the satellite spots disappear, but retains the 1st Bi BL LEED spots, showing the coexistence of the single Bi BL and Bi(0001) bulk patterns. Also, the corresponding TOF spectrum in the middle panel of Fig. 9 doesn't show any significant changes in the intensity of either the Se or Bi SSP, indicating that Bi is not removed in a manner that reveals the substrate. By measuring the distance between the LEED spots (Fig. 8(b), lines E and F), corresponding lattice constants of 4.20 Å and 4.57 Å are found, which have a relative difference of about 9%. This can be interpreted as the 2$^{nd}$ and 3$^{rd}$ Bi BLs collapsing under annealing and forming bulk-like Bi single crystal islands, while the 1$^{st}$ Bi bilayer remains stable and covers the substrate as a commensurate layer. The AFM images in Figs. 1(f)-(g) confirm this assumption. Figure 1(f) was collected from a Bi$_2$Se$_3$ surface covered with 3 Bi BLs and annealed at 120°C for 10 hours, and it shows the appearance of holes and islands. The red line C in Fig. 1(g) shows the line profile from Fig. 1(f), which displays island heights up to 5 BLs and 1 BL deep holes. As the total coverage is 3 Bi BLs, it can be inferred that there is a single Bi BL covering the substrate at the bottom of the holes, while the 2$^{nd}$ and 3$^{rd}$ Bi BLs crystalize to form islands with heights of 4 or 5 BLs.

When the annealing temperature is 490°C, the surface of the sample reverts to the stable Se-terminated Bi$_2$Se$_3$ surface produced by IBA, as shown by the LEED pattern (Fig. 8(c)) and TOF spectra (bottom panel of Fig. 9). This indicates that the Bi film sublimates, perhaps along with some substrate QLs, and the surface recrystallizes to produce a fresh Se-terminated surface, as occurs after annealing in the IBA process.

## IV. Discussion

The measurements as a function of Bi deposition are used to determine the growth mechanism of Bi films on Bi$_2$Se$_3$ surfaces. A combination of surface analysis tools shows that the first BL grows flat while subsequent layers initially form thin "islands". This is similar to the Stranski–Krastanov (SK) growth mode, indicating a change in the surface energy with film thickness. For Bi/Bi$_2$Se$_3$, however, these "islands" are no higher than 2 BLs and coalesce into complete bilayers before thicker islands build up, which is because the bonding between the layers in 2D materials, such as Bi, is relatively weak. Therefore, the growth is not purely SK, and is thus often referred to in the literature as being a quasi bilayer-by-bilayer mode, such as in Refs. [20,49]. These references also observe that Bi grows in a step-flow mode at higher temperature due to the increased surface mobility of Bi adatoms, but the growth of Bi at room temperature is at the boundary of the transition between these two modes.

The first bilayer grows commensurately on fcc-sites of the Se-terminated Bi$_2$Se$_3$ surface with a small domain size. As shown in Fig. 5(b), if the Bi bilayer sits on fcc-sites of the substrate, the structure of the top layers is basically the same as pristine Bi$_2$Se$_3$, except that the first Se layer is replaced by Bi atoms. Therefore, the ICISS scans maintain the features of the original surfaces, except for the loss of the Se surface flux peak. Meanwhile, a simulation of the ICISS scans using a superlattice, in which both the adlayer and substrate maintain their own lattice parameters, does not agree with the experimental ICISS data. Since the estimated domain size from LEED for 1 BL of deposited Bi is around 19 atoms, it's reasonable to infer that the surface does not consist of large superlattices, but instead of small domains in which the Bi atoms are located at fcc-like sites above the Bi$_2$Se$_3$ substrate and the strain between the Bi bilayer and substrate doesn't accumulate over a long range. The lattice parameter measured from the LEED patterns and in Ref. [19] is 4.2 Å, a bit larger than the substrate parameter of 4.14 Å, which may indicate some lateral expansion of the local domains. Thus, the deposited Bi atoms favor forming small domains instead of a superlattice.

When the 2$^{nd}$ bilayer grows, its lattice constant jumps up from 4.20 Å to 4.48 Å, which is a lattice mismatch of 6.7%. Recent work using scanning tunneling microscopy (STM) shows that Bi BLs epitaxially grown on $Bi_2Se_3$ have a similar large lattice mismatch of 3.3% between the 1st and 2nd Bi BLs [19]. The small difference in the measured lattice mismatch numbers between the present results and those of Ref. [19] are possibly due to the nature of the techniques. STM probes the well-ordered portions of the sample locally, while LEED averages over a large area of the sample surface.

For coverages from 2 to 6 BL, satellite spots appear in the LEED images, indicating that a Moiré structure forms on substrate. The appearance of a Moiré structure can, in principle, be caused by two different scattering mechanisms [50]:

1. Multiple scattering between a substrate and film with different lattice parameters.
2. An undulation of a film that is locked in at nearly commensurate substrate positions, which forms a phase grating for the scattered electrons.

The satellite spots show up when the 2nd Bi BL is deposited and remain until the substrate is covered by 8 BLs. Note that the LEED detection depth is about 5-10 Å [51], which is close to the thickness of 2 Bi BLs. Therefore, the persistence of the satellite spots in LEED indicates that it is most likely due to a periodic undulation of the surface. This conclusion is consistent with the STM study of Ref. [19], where a periodic corrugation is observed for $Bi_2Se_3$ samples covered by 2 to 5 Bi BLs, which attenuates after more Bi is deposited.

The corrugation between the 1st and subsequent BLs releases the strain caused by the large lattice mismatch of the different materials. $Bi_2Se_3$ has lattice constant of 4.14 Å while bulk Bi(0001) has lattice constant of 4.54 Å, which is a mismatch of 9.6%. The lattice mismatch is accommodated between the 1st and 2nd Bi BL, rather than between the surface Se and the 1st Bi BL, which is consistent with there being a stronger bond between the top layer Se in the substrate and the 1st Bi BL than between the additional Bi BLs.

Evidence for the relatively strong bond between the 1st Bi BL and the substrate is supported by several of the measurements performed here. After annealing at 120°C, for example, the 1st BL stays in place, as shown by the strong Bi SSP in the double alignment TOF-LEIS spectrum in the center panel of Fig. 9. The AFM image in Fig. 1(f)) shows that the additional Bi diffuses to form islands, which is a lower energy configuration. The formation of islands releases the strain in the 2nd Bi BL caused by the large lattice mismatch between the 1st and 2nd Bi BL, while the 1st BL Bi doesn't diffuse away due to the stronger bonds between Bi and the surface Se in $Bi_2Se_3$. The stronger bonding of the 1st BL to the substrate is also indicated by the other AFM images. The 1st Bi BL islands on bare $Bi_2Se_3$ (Fig. 1(b)) and the islands created from the 2nd Bi BL (Fig. 1(d)) have different shapes. The single Bi BL islands are more stretched and flat while the 2nd bilayer islands are more triangular shaped. This implies that the bonding between $Bi_2Se_3$ and a single Bi BL is different than that between Bi BLs. All of this provides evidence of the particular stability of a single BL-covered $Bi_2Se_3$ surface, consistent with calculations in the literature [21-24].

## V. Conclusions

The deposition of Bi BLs on $Bi_2Se_3$(0001) is studied by LEED, LEIS and AFM as a function of coverage and annealing temperature. The first Bi bilayer is locally commensurate, confirming results from the literature that employed different experimental techniques [19]. Furthermore, it is shown here that the BL adsorbs on fcc-sites of the $Bi_2Se_3$ substrate with an average domain size that is less than 20 atoms in diameter. Charge transfer from the first BL to the substrate is demonstrated by the neutralization of scattered $Na^+$ ions, and this charge transfer is different for the outermost and second layer Bi atoms in that 1st BL. Satellite spots appear when the 2nd Bi BL is deposited indicating a corrugation of the 2nd BL. Additional deposited Bi grows as free-standing bilayers that follow the corrugation. The outermost surface eventually flattens to become a film with the structure of bulk Bi when more than 8 BLs are deposited. Annealing leaves the 1$^{st}$ Bi BL intact, while the additional Bi diffuses to form taller islands. This indicates a strong interaction between the substrate $Bi_2Se_3$ QL and the 1$^{st}$ Bi BL, which is consistent with theoretical predictions and other experimental results that suggest that a single Bi BL-covered $Bi_2Se_3$ surface is a particular stable structure. This explains why Bi-rich surfaces have been observed to be stable following cleaving under certain conditions [5,22,25,26]. The interaction between Bi BLs is not as strong, which leads to the formation of islands of bulk Bi following annealing of samples with larger coverages.

## VI. Acknowledgements

This material is based on work supported by the U.S. Army Research Laboratory and the U.S. Army Research Office under Grant No. 63852-PH-H.

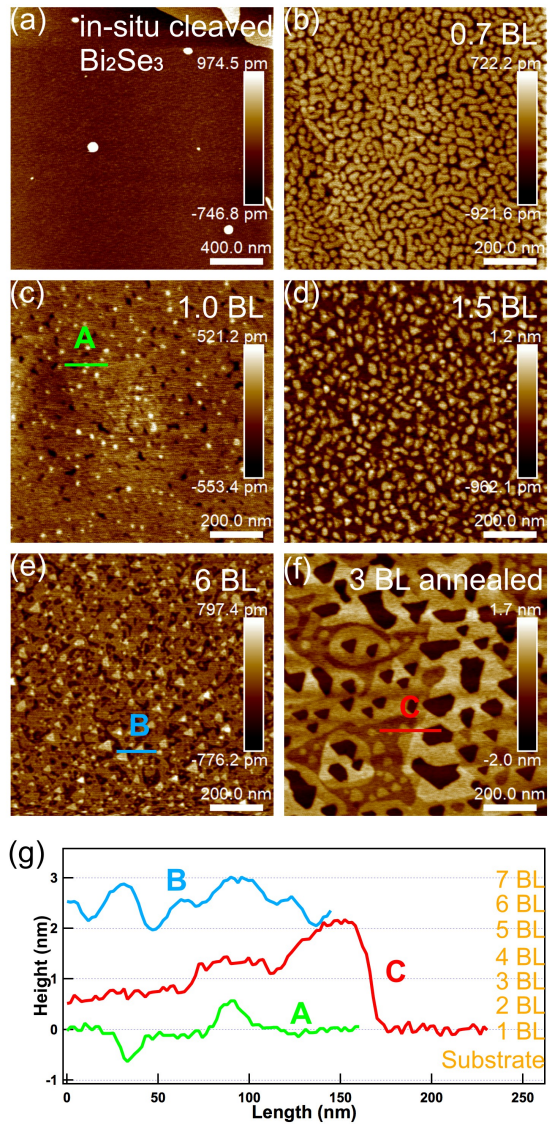

**Figure 1.** AFM images collected from (a) an *in-situ* cleaved Bi$_2$Se$_3$ surface, and from Bi films grown on Bi$_2$Se$_3$ with evaporation times of (b) 2.5 min, (c) 3.5 min (d) 5.5 min and (e) 21 min, which lead to the indicated Bi coverages (see text), (f) 3 Bi BLs grown on Bi$_2$Se$_3$ and annealed at 120°C for 10 hrs, (g) height profiles of lines A, B and C that are drawn in (c), (d) and (e).

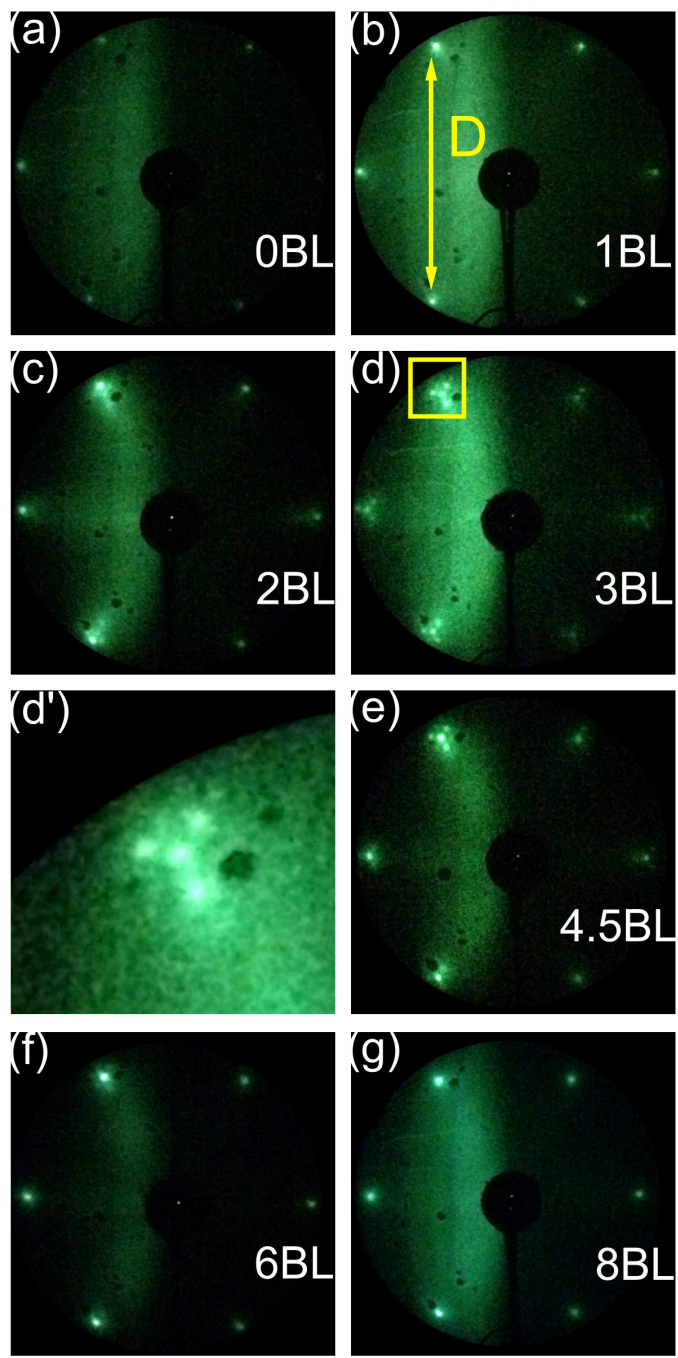

**Figure 2.** LEED patterns collected from $Bi_2Se_3$ surfaces with the indicated Bi coverages. (d') is a close-up of the boxed area in (d). All of the LEED images were collected using a beam energy of 21.8 eV.

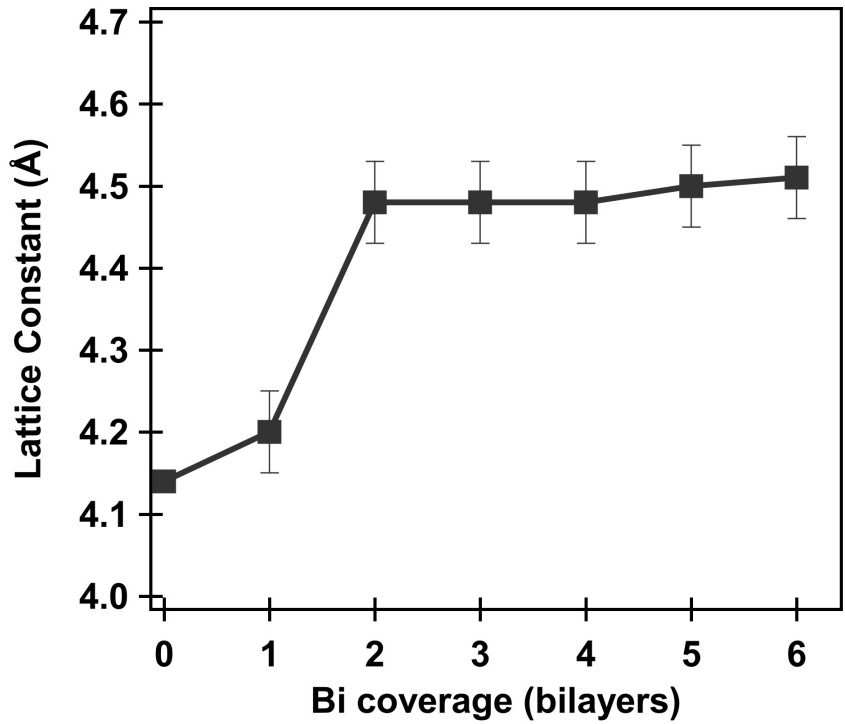

**Figure 3.** Top layer lattice constant as a function of Bi coverage calculated using the length of line D in Fig. 2(b) and calibrated with the known $Bi_2Se_3$ lattice constant of 4.14 Å.

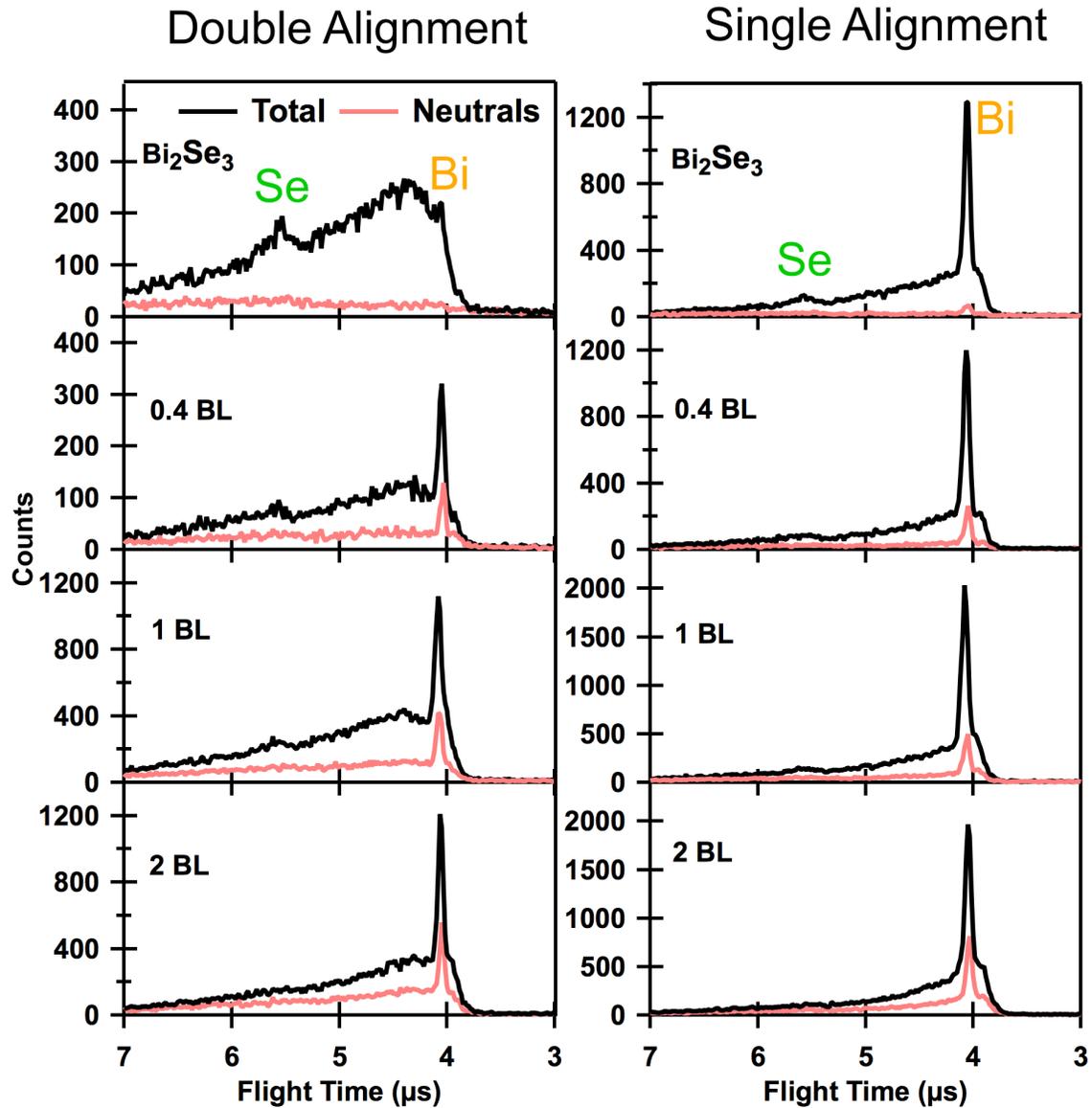

**Figure 4.** LEIS TOF spectra collected from $Bi_2Se_3$ with the indicated Bi coverages. A 3.0 keV $Na^+$ ion beam is incident normal to sample surface and the detector is positioned at a scattering angle of 125°. In each spectrum, the black line represents total yield and red line represents the scattered neutral particles. The spectra are normalized to each other using the ion beam current measured on the sample and the data collection time so that they can be quantitatively compared.

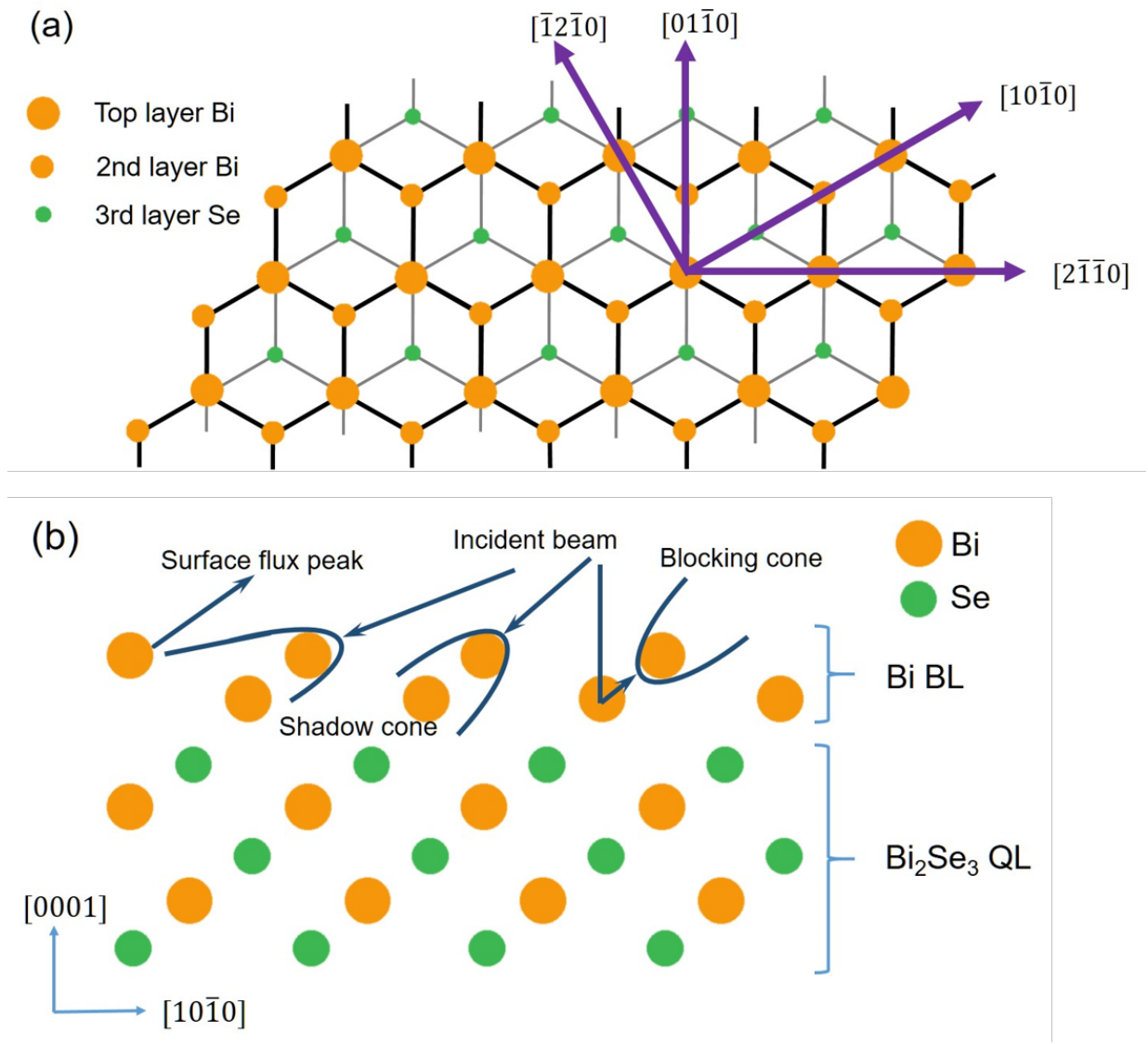

**Figure 5.** (a) Top view of a single Bi bilayer on a Bi$_2$Se$_3$(0001) surface. (b) Side view of the (1$\bar{2}$10) plane of 1 Bi bilayer on Bi$_2$Se$_3$(0001). Representative shadow and blocking cones are indicated in (b).

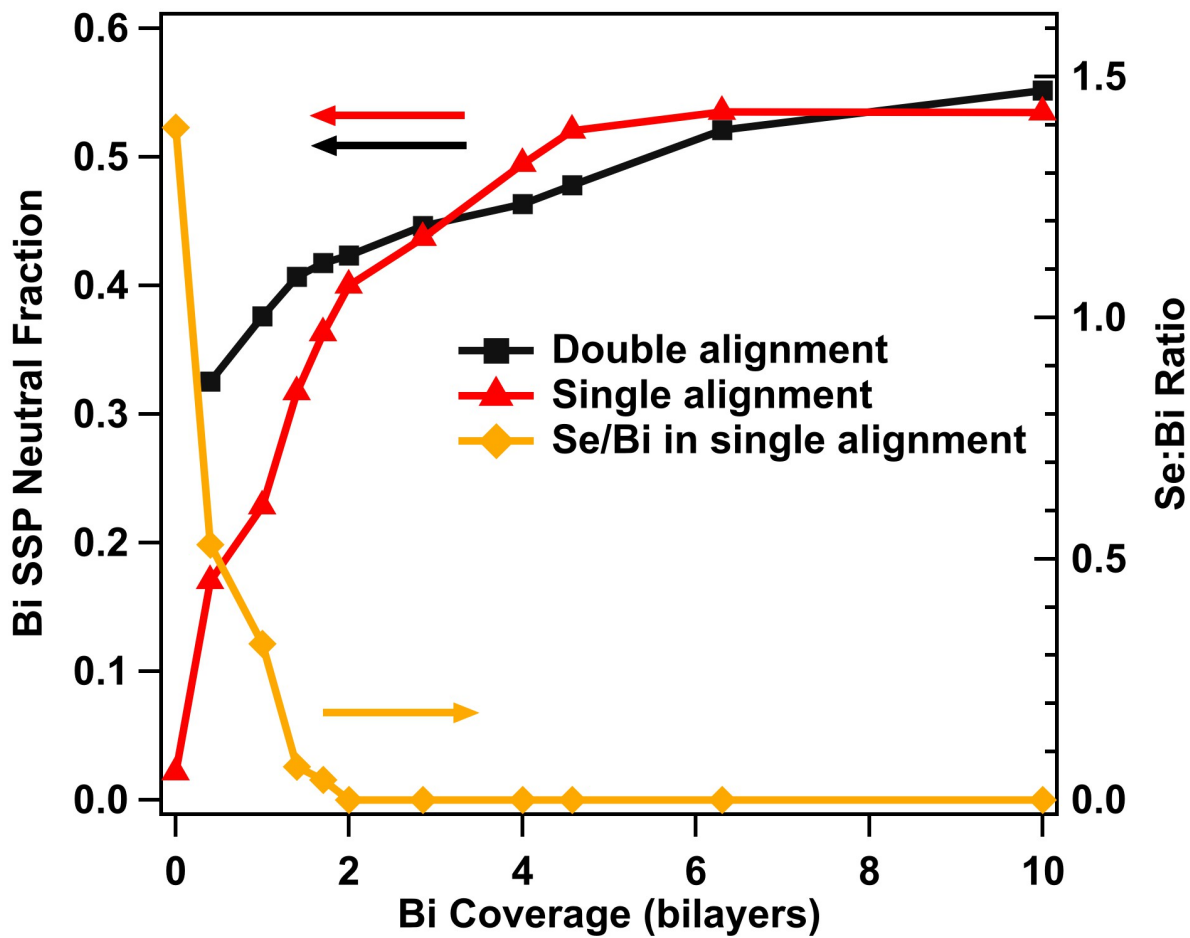

**Figure 6.** Neutral fractions of the Bi SSP for 3.0 keV Na$^+$ scattered from Bi$_2$Se$_3$ in both single and double alignment along with the Se:Bi ratio calculated from ion scattering data collected in single alignment, shown as a function of Bi coverage.

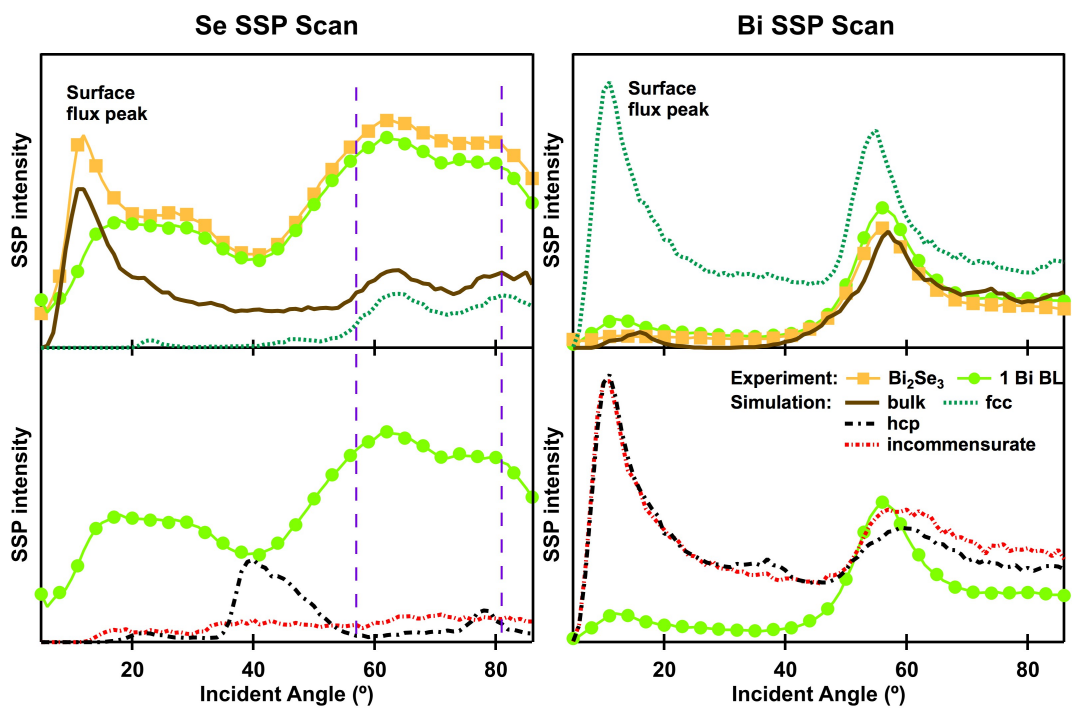

**Figure 7.** Impact collision ion scattering spectroscopy (ICISS) polar angle scans and simulations for the Se (left panels) and Bi (right panels) SSPs collected from $Bi_2Se_3$ and the 1 Bi BL covered surface using 3.0 keV $Na^+$ ions at a scattering angle of 161°. The sample rotates in the polar direction along the $[\bar{1}010]$ azimuth so that the ion incidence direction gradually varies from the surface plane to the surface normal. The surface flux peaks are indicated in upper panels. The legend in bottom right panel applies to all of the graphs.

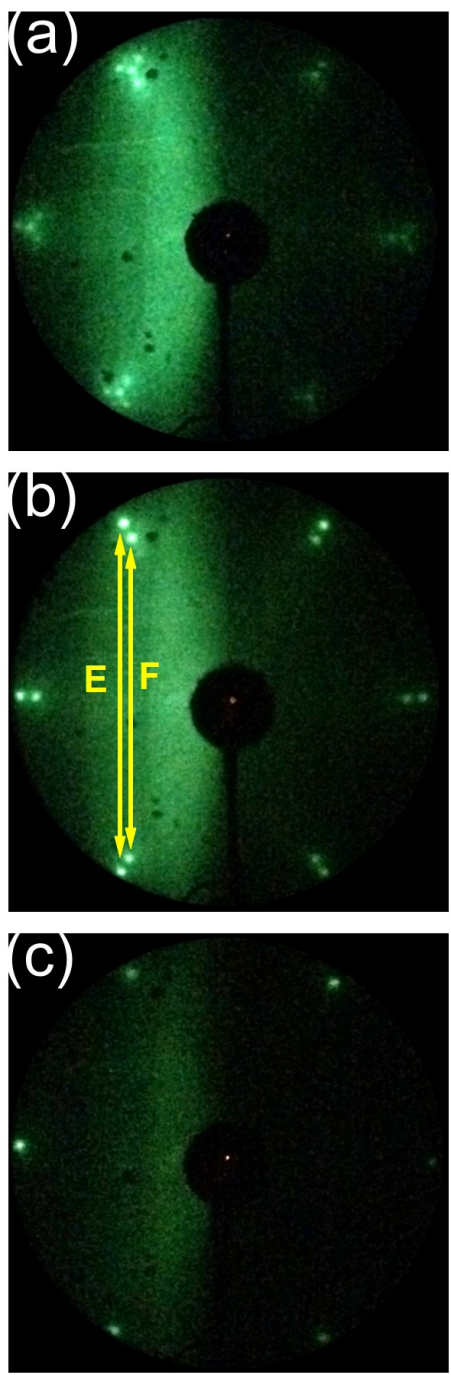

**Figure 8.** LEED patterns collected from 3 Bi BLs deposited on $Bi_2Se_3$ (a) before annealing, (b) after annealing at 120°C for 10 hrs, and (c) after annealing at 490°C for 0.5 hrs.

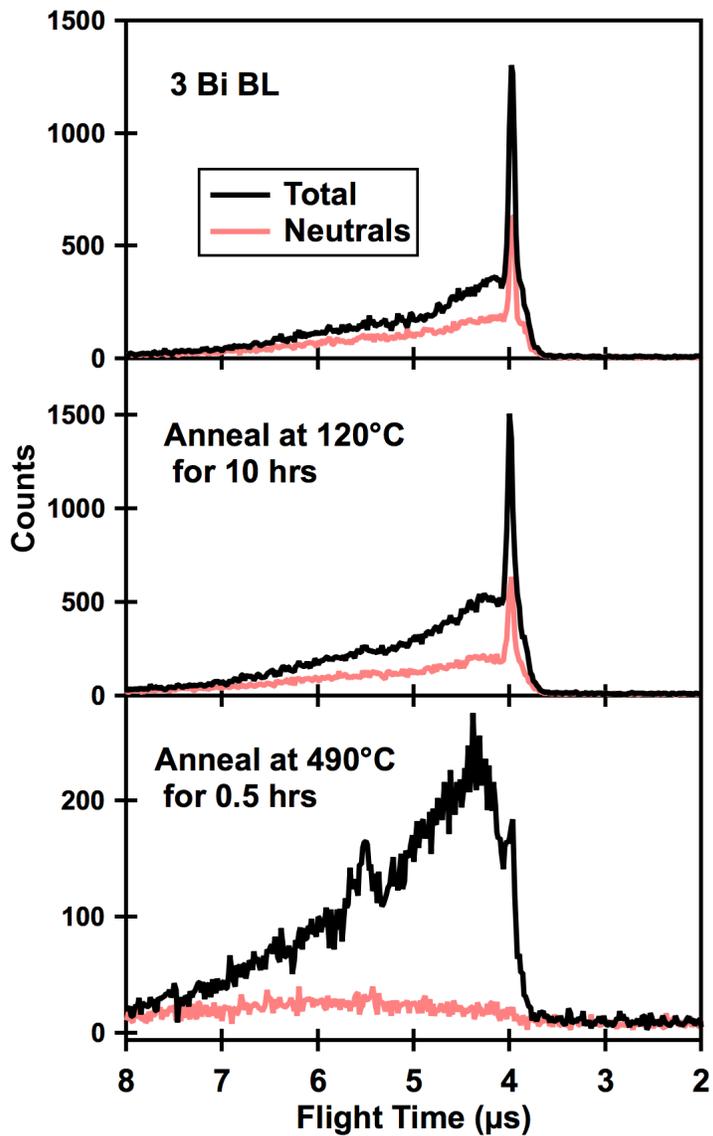

**Figure 9.** LEIS TOF double alignment spectra collected using 3.0 keV Na$^+$ ions scattered from 3 Bi BLs deposited on a Bi$_2$Se$_3$(0001) surface and after annealing at the indicated temperatures for the indicated times.